\documentclass{iau} 
\usepackage{graphicx}
\usepackage{natbib}
\usepackage{hyperref}
\usepackage{siunitx}

\newcommand{\GG}{\mbox{$G$}}
\newcommand{\GBP}{\mbox{$G_{\rm BP}$}}
\newcommand{\GRP}{\mbox{$G_{\rm RP}$}}
\newcommand{\Zthree}{\mbox{$Z_{\rm EDR3}$}}
\newcommand{\pic}{\mbox{$\varpi_{\rm c}$}}
\newcommand{\sigmas}{\mbox{$\sigma_{\rm s}$}}
\newcommand{\sigmai}{\mbox{$\sigma_{\rm int}$}}
\newcommand{\sigmae}{\mbox{$\sigma_{\rm ext}$}}
\newcommand{\microas}{\mbox{$\mu$as}}
\newcommand{\lili}{\mbox{LiLiMaRlin}}
\newcommand{\VO}[1]{Villafranca~O-{#1}}

\title[The Gaia view on massive stars] 
{The Gaia view on massive stars: \linebreak EDR3 and what to expect from DR3}

\author[Jes\'us Ma\'{\i}z Apell\'aniz \textit{et al.}]   
{Jes\'us Ma\'{\i}z Apell\'aniz$^{1}$,
 Rodolfo H. Barb\'a$^{2}$\footnote{Deceased},
 Michelangelo Pantaleoni Gonz\'alez$^{1,3}$,
 Michael Weiler$^{4}$,
 B. Cameron Reed$^{5}$,
 Rom\'an Fern\'andez Aranda$^{1,3,6}$,
 Pablo Crespo Bellido$^{1,3}$,
 Alfredo Sota$^{7}$,
 Emilio J. Alfaro$^{7}$,
 \and J. Alejo Molina Lera$^{1,8}$}

\affiliation{$^1$Centro de Astrobiolog\'{\i}a (CAB), CSIC-INTA. Campus ESAC. E-\num[detect-all]{28692} Madrid, Spain.
\\[\affilskip]
$^{2}$Dept. de Astronom\'{\i}a, Universidad de La Serena. Av. Cisternas 1200 Norte. La Serena, Chile.
\\[\affilskip]
$^3$Dept. de Astrof{\'\i}sica y F{\'\i}sica de la Atm\'osfera, U. Compl. de Madrid. E-\num[detect-all]{28040} Madrid, Spain.
\\[\affilskip]
$^4$Dept. de F{\'\i}sica Qu\`antica i Astrof{\'\i}sica, U. de Barcelona (IEEC-UB). 08\,028. Barcelona, Spain.
\\[\affilskip]
$^5$Department of Physics (Emeritus), Alma College. \num[detect-all]{48801} Michigan. United States of America.
\\[\affilskip]
$^6$Institute of Astrophysics, University of Crete. \num[detect-all]{70013} Heraklion, Greece.
\\[\affilskip]
$^7$Instituto de Astrof\'{\i}sica de Andaluc\'{\i}a (IAA), CSIC. E-\num[detect-all]{18008} Granada, Spain.
\\[\affilskip]
$^8$Instituto de Astrof\'{\i}sica de La Plata (CONICET, UNLP). 1900 La Plata, Argentina.
}

\pubyear{2022}
\volume{361}  
\jname{Massive Stars Near and Far}
\editors{N.~St-Louis, J.~S.~Vink \& J.~Mackey, eds.}
\begin{document}

\maketitle

\begin{abstract}
At the time of this meeting, the latest {\it Gaia} data release is EDR3, published on 3 December 2020, but the next one, DR3, will appear soon, on 13 June 2022.
This contribution describes, on the one hand, {\it Gaia}~EDR3 results on massive stars and young stellar clusters, placing special emphasis on how a correct
treatment of the astrometric and photometric calibration yields results that are simultaneously precise and accurate. On the other hand, it gives a brief description
of the exciting results we can expect from {\it Gaia}~DR3.
\end{abstract}

\firstsection 
\section{Introduction}

The early third {\it Gaia} data release (EDR3, \citealt{Browetal21}) provides astrometry for \num{1467744819} sources and \GG-band photometry for \num{1806254432}
sources. For the astrometry, 40\% of the sources are ``five-parameter solutions'' in which the (pseudo-)color of the source is known independently and is of higher 
quality than the remaining 60\% of ``six-parameter solutions'' in which the (pseudo-)color has to be simultaneously calculated \citep{Lindetal21a}. For the
photometry, in addition to the \GG-band information, 86\% of the sources also have \GBP\ and/or \GRP\ magnitudes extracted from the spectrophotometric detectors
\citep{Rieletal21}, whose wavelength-dependent information will not be available until the full third data release (DR3).

This contribution has four parts. First, we describe our calibration efforts to ensure the precision and accuracy of the astrometry
and photometry for all sources, massive or not. Second, we present the integration between {\it Gaia} data and the Galactic OB stars of the Alma Luminous Star 
(ALS) catalog \citep{Reed03}, started with the previous {\it Gaia} data release (DR2, see \citealt{Pantetal21}) and which we have now updated with EDR3
data to produce the largest catalog of massive Galactic OB stars with accurate distances ever. Third, we discuss the
Villafranca project of Galactic young stellar groups, which is cataloguing all stellar groups with OB stars in the solar neighborhood 
\citep{Maiz19,Maizetal20b,Maizetal22a}. Finally, we briefly comment on what to expect from the new data products that will become available in DR3.


\section{Astrometric and photometric calibration}

\citet{Lindetal21b} describe that the catalog EDR3 parallaxes $\varpi$ have a zero point \Zthree\ that depends on magnitude,
color, and ecliptic latitude and that to minimize systematic effects one should use a corrected parallax $\pic = \varpi - \Zthree$. In two papers 
\citep{Maizetal21c,Maiz22} we have built on that work to produce a methodology that yields accurate and precise EDR3 parallaxes:

\begin{itemize}
 \item \citet{Maiz22} provides an alternative \Zthree\ with the same functional form as \citet{Lindetal21b}. The parameters are very similar for $\GG>13$ 
       but differ significantly for brighter stars.
 \item As described in \citet{Maizetal21c} and elsewhere (e.g. \citealt{Fabretal21a}), EDR3 parallax uncertainties \sigmai\ are too low and, if used
       directly, yield distances with underestimated uncertainties. Following the analysis of \citet{Maiz22}, calculating external uncertainties \sigmae\ as
       $\sigmae = \sqrt{k^2 \sigma_{\rm int}^2 + \sigma_{\rm s}^2}$, where $k$ is a magnitude-dependent multiplicative constant and \sigmas\ is described below, 
       provides more realistic uncertainty estimates. $k$ is close to 1 for the fainter stars, increases to values close to 2 around $\GG = 12$, decreases again to 
       $\sim$1.3 around $\GG = 10$, and can become as large as 3.0 for the brightest stars, where the original \Zthree\ was based on a small number of sources.
 \item EDR3 parallaxes are correlated as a result of the presence of a residual angular covariance \citep{Lindetal21a}. This has two consequences. [1] The existence
       of an uncorrected systematic uncertainty for individual parallaxes (\sigmas\ in the previous formula) that is the square root of the angular covariance at 
       zero separation. It has an estimated value of 10.3~\microas\ and can be interpreted as the minimum uncertainty of any single EDR3 parallax \citep{Maizetal21c}. 
       [2] The need to add covariance terms when combining parallax uncertainties of objects assumed to be at the same distance (e.g. stellar clusters). As a corollary,
       stellar clusters that span a small angle in the sky have combined parallax uncertainties that approach \sigmas\ asymptotically, leading to minimum 
       distance uncertainties of $\sim$1\% at 1~kpc and $\sim$3\% at 3~kpc \citep{Maizetal22a}.
 \item Parallaxes for objects with six-parameter solutions or large RUWE (Renormalized Unit Weight Error) are, in general, of worse quality but it is possible to use 
       them correctly with an increased value of $k$ \citep{Maiz22}.
\end{itemize}

With respect to photometry, {\it Gaia} holds the promise of providing the first high dynamic range (17 mag), simultaneously accurate and precise (better that 
10~mmag), whole-sky multiband optical photometric survey. However, as we discovered for {\it Gaia}~DR1 \citep{Maiz17a} and DR2 \citep{MaizWeil18}, eliminating the
residual systematic offsets requires tweaking the filter passbands and applying small corrections. We have produced a similar analysis for EDR3 that will be 
submitted soon (Weiler et al. in preparation).

\section{The ALS catalog and distances to OB stars}

Finding OB stars via photometry alone, though possible \citep{MaizSota08,Maizetal14a}, is difficult due to the combined effects of extinction and photometric
calibration of ground-based surveys. Two decades ago, \cite{Reed03} used a combination of photometric and (preferably) spectroscopic data to compile the ALS catalog
of Galactic OB stars, which by 2005 contained \num{18693} objects. However, the diverse quality of the data (which in many cases lacked even good-quality
coordinates) hampered the use of the catalog without including a significant number of contaminants. 

In \citet{Pantetal21}, from now on ALS~II, we undertook the task of cross-matching the ALS catalog with {\it Gaia}~DR2, painstakingly comparing star by star between
the existing data and the {\it Gaia} catalog. After eliminating duplicates, unmatched objects, and stars with bad astrometry or photometry, we were left with
\num{15662}, of which \num{13762} are suspected of truly being Galactic OB massive stars. Among the results from that paper, we point out the following:

\begin{itemize}
 \item Neither Apsis \citep{Bailetal13} nor StarHorse \citep{Andeetal19} provide good extinction measurements for OB stars based on {\it Gaia}~DR2 data. Apsis apparently 
       mistakes high-extinction OB stars for cooler, lower-extinction objects while StarHorse allows for some objects to have large negative extinctions and others to 
       follow an anomalous relationship between color excess and amount of extinction.
 \item In order to convert parallaxes into distances one needs to use priors. Commonly used with {\it Gaia}~DR2 data are those of \citet{Bailetal18} and
       \citet{Andeetal19}, which assume spatial distributions different from the one for OB stars. For that reason, we compared their distances with the ones derived
       from the prior of \citet{Maiz01a,Maiz05c} and the updated parameters of \citet{Maizetal08a}. We found good agreement among the distances derived from the
       three priors for values below 3~kpc but some differences beyond that. 
 \item We produced a map of the location of the OB stars in the solar neighborhood. The most significant result is the discovery of a structure that extends from the
       Orion-Cygnus (or Local) and Perseus arms that we dub the Cepheus spur. It is located 50-100~pc above the Galactic plane and appears to be a 2-D extension of
       the 1-D Radcliffe wave \citep{Dixo67,Alveetal20}.
\end{itemize}

After our work with ALS~II, {\it Gaia}~EDR3 yielded significantly improved astrometry and photometry. We continued our analysis of the data by attempting to
cross-match the ALS stars that were not in the \num{15662} sample above and reprocessing the whole sample to better differentiate between massive and non-massive
stars. We have also expanded the sample by including stars with GOSSS \citet{Maizetal11} and \lili\ \citep{Maizetal19a} spectral classifications (a new paper of the
GOSSS series, GOSSS~IV, will be submitted soon) and from the Cygnus~OB2 \citep{Berletal20} and Carina~OB1 (Berlanas et al. in
preparation) associations. The new sample includes several thousand more stars and will be submitted as ALS~III later this year. 

The information that will be made available in DR3 (see below) and that from ground-based surveys such as GALANTE \citep{Maizetal21d} will allow us to
significantly expand the ALS sample in the near future. Our estimate is that as many as $10^5$ Galactic OB stars may be identified using those data.



\section{The Villafranca project}

The Villafranca catalog of Galactic OB stellar groups combines {\it Gaia} astrometry and photometry with spectroscopy from ground-based surveys to derive information about 
the young stellar clusters and associations in the solar neighborhood and beyond. A zeroth paper \citep{Maiz19} describes the method used to determine group
membership, Villafranca~I \citep{Maizetal20b} analyzes 16 groups with O2-O3.5 stars using {\it Gaia}~DR2 data, and Villafranca~II \citep{Maizetal22a} revises those results 
with {\it Gaia}~EDR3 data and adds ten new stellar groups with O stars.

With the Villafranca project we have been able to provide precise distances ($\sim$1\% at 1~kpc, $\sim$3\% at 3~kpc) whose accuracy has been verified for some examples using
alternative geometric distances. We have also published detailed analyses of each cluster, describing the massive stars with accurate spectral types that belong to them and
finding runaway candidates. For four nearby low-extinction clusters we have used the PMS seen in the cleaned {\it Gaia} CMDs to estimate their ages. Such CMD analyses will be
extended in the future with the help of complementary ground-based surveys such as GALANTE \citep{Maizetal21d}.

The most interesting result that has come out of the Villafranca project is the discovery of orphan clusters, very young stellar groups that have been disrupted as the
consequence of multiple stellar interactions among their massive stars. The prototype object is the Bermuda cluster (\VO{014}~NW) at the North~America nebula
\citep{Maizetal22b}, which through three ejection events expelled a total of at least nine stellar systems (twelve individual stars) between 1.9 and 1.5~Ma ago, including
its three most massive stars (one of which, the primary of the Bajamar system, is the earliest-type O star within 1~kpc of the Sun). As a result, the cluster has lost over
200~M$_\odot$ in stars and has altered its mass function from one that was top-heavy to one that is compatible with a Kroupa-like function. Moreover, the cluster has lost so 
much mass that it is now expanding \citep{Kuhnetal20} and will likely dissolve several million years from now.

We are currently working on the third paper of the Villafranca series, which will concentrate on the region of the sky where Carina~OB1\footnote{Which, despite its name, 
should not be considered the Keel Kenobi association, something that was instead a result of Darth Vader's actions.} is located. Three stellar groups in Carina~OB1 
(\VO{002}, \VO{003}, and \VO{025}) and one in its background (\VO{004}) were already analyzed in previous papers and here we will add several more, some with O and early-B 
stars and some with just the latter. Future papers will expand the sample of Villafranca groups.

\section{A peek into what the future holds in store}

By the time this contribution is published, {\it Gaia}~DR3 will be out in the wild but at the time of this writing access to the data is restricted to DPAC members (and
their lips are sealed). What can we say about it without making fools of ourselves in the eyes of you, dear (future) reader, who may know more than us?

We can divide the DR3 results into two types: {\it raw} and {\it processed}. By {\it raw} we mean the direct output of the {\it Gaia} instruments without significant 
assumptions about the nature of the source and/or comparisons with models and by {\it processed}
those results that imply an additional treatment such as measuring velocities, equivalent widths, or time
series of the raw data. There will be three types of raw results: low-resolution BP+RP spectrophotometry for \num{219197643} objects, high-resolution $z$-band spectroscopy
for \num{999645} objects, and $z$-band photometry for \num{32232187} objects, in all cases averaged over time. The list of processed results is significantly longer and
includes radial velocity and $v\sin i$ measurements, variable-star identifications and classifications, astrometric and spectroscopic orbit determinations, astrophysical
parameters estimations, and H$\alpha$ and DIB~$\lambda$8621 equivalent widths, among others.

The information content resulting from {\it Gaia}~DR3 will surpass that of previous data releases, as two dimensions will be added to a large number of objects: wavelength
in the form of the two types of spectroscopy and time in the form of the analysis of variable stars. It is true that, as it was before, most of the information
will not be for massive stars but this time there has been a concerted effort by DPAC to find hot stars (mostly of intermediate mass but some massive) in the sample through 
a combination of the low-resolution spectrophotometry and the high-resolution spectroscopy. Furthermore, the added information will surely help finding new massive needles
in the Galactic haystack.

Further down the road we will have {\it Gaia}~DR4. While EDR3 and DR3 are based on 34 months of data, DR4 will use almost double that amount, 66 months. Such a longer
baseline will undoubtedly improve the precision of most data products, especially proper motions. However, an even better improvement is likely to take place in the
accuracy, that is, the reduction of systematic errors, as the experience from previous data releases will be used to characterize and correct them (for example, the
comparison between DR2 and EDR3 was the base for the work of \citealt{Lindetal21b}, see section~3 in that paper). And then there will be the data deluge in the form of the
full astrometric, photometric, and radial-velocity catalogs: all data, all epochs, enough to maintain a generation of Galactic astronomers busy.

\section*{Acknowledgements}
One of the authors, Rodolfo Barb\'a unexpectedly passed away after making significant contributions to most aspects of this work: we dedicate it to him.
J.M.A., M.P.G., R.F.A., P.C.B., A.S., and J.A.M.L. acknowledge support from the Spanish Government Ministerio de Ciencia e Innovaci\'on through grant PGC2018-095\,049-B-C22. 
M.W. acknowledges funding by the Spanish MICIN/AEI/\num{10.13039}/\num{501100011033}, by ``ERDF A way of making Europe'' by the ``European Union'' through grant 
RTI2018-095\,076-B-C21, and by the Institute of Cosmos Sciences University of Barcelona (ICCUB, Unidad de Excelencia ’Mar\'{\i}a de Maeztu’) through grant CEX2019-000\,918-M.
E.J.A. acknowledges support from the State Agency for Research of the Spanish Government Ministerio de Ciencia e Innovaci\'on through the 
``Center of Excellence Severo Ochoa'' award to the Instituto de Astrof{\'\i}sica de Andaluc\'{\i}a (SEV-2017-0709) and through grant PGC2018-\num{095049}-B-C21
This work has made use of data from the European Space Agency (ESA) mission {\it Gaia}\footnote{\url{https://www.cosmos.esa.int/gaia}}, 
processed by the {\it Gaia} Data Processing and Analysis Consortium (DPAC\footnote{\url{https://www.cosmos.esa.int/web/gaia/dpac/consortium}}).
Funding for the DPAC has been provided by national institutions, in particular the institutions participating in the {\it Gaia} 
Multilateral Agreement. 
The {\it Gaia} data is processed with the computer resources at Mare Nostrum and the technical support provided by BSC-CNS.

\bibliographystyle{apj}
\bibliography{general}

\begin{thebibliography}{}
\expandafter\ifx\csname natexlab\endcsname\relax\def\natexlab#1{#1}\fi

\bibitem[{Alves {et~al.}(2020)Alves, Zucker, Goodman, Speagle, Meingast,
  Robitaille, Finkbeiner, Schlafly, \& Green}]{Alveetal20}
Alves, J., Zucker, C., Goodman, A.~A., {et~al.} 2020, Nature, 578, 237

\bibitem[{Anders {et~al.}(2019)Anders, Khalatyan, Chiappini, Queiroz, Santiago,
  Jordi, Girardi, Brown, Matijevi{\v{c}}, Monari, Cantat-Gaudin, Weiler, Khan,
  Miglio, Carrillo, Romero-G{\'o}mez, Minchev, de~Jong, Antoja, Ramos,
  Steinmetz, \& Enke}]{Andeetal19}
Anders, F., Khalatyan, A., Chiappini, C., {et~al.} 2019, A\&A, 628, A94

\bibitem[{Bailer-Jones {et~al.}(2018)Bailer-Jones, Rybizki, Fouesneau,
  Mantelet, \& Andrae}]{Bailetal18}
Bailer-Jones, C.~A.~L., Rybizki, J., Fouesneau, M., Mantelet, G., \& Andrae, R.
  2018, AJ, 156, 58

\bibitem[{Bailer-Jones {et~al.}(2013)Bailer-Jones, Andrae, Arcay, Astraatmadja,
  Bellas-Velidis, Berihuete, Bijaoui, Carri{\'o}n, Dafonte, Damerdji,
  Dapergolas, de~Laverny, Delchambre, Drazinos, Drimmel, Fr{\'e}mat, Fustes,
  Garc{\'\i}a-Torres, Gu{\'e}d{\'e}, Heiter, Janotto, Karampelas, Kim, Knude,
  Kolka, Kontizas, Kontizas, Korn, Lanzafame, Lebreton, Lindstr{\o}m, Liu,
  Livanou, Lobel, Manteiga, Martayan, Ordenovic, Pichon, Recio-Blanco,
  Rocca-Volmerange, Sarro, Smith, Sordo, Soubiran, Surdej, Th{\'e}venin,
  Tsalmantza, Vallenari, \& Zorec}]{Bailetal13}
Bailer-Jones, C.~A.~L., Andrae, R., Arcay, B., {et~al.} 2013, A\&A, 559, A74

\bibitem[{Berlanas {et~al.}(2020)Berlanas, Herrero, Comer{\'o}n,
  Sim{\'o}n-D{\'\i}az, Lennon, Pasquali, Ma{\'\i}z~Apell{\'a}niz, Sota, \&
  Peller{\'\i}n}]{Berletal20}
Berlanas, S.~R., Herrero, A., Comer{\'o}n, F., {et~al.} 2020, arXiv e-prints,
  arXiv:2008.09917

\bibitem[{Brown {et~al.}(2021)Brown, Vallenari, Prusti, de~Bruijne, Babusiaux,
  Biermann, Creevey, Evans, Eyer, \& et~al.}]{Browetal21}
Brown, A.~G.~A., Vallenari, A., Prusti, T., {et~al.} 2021, A\&A, 649, A1

\bibitem[{Dixon(1967)}]{Dixo67}
Dixon, M.~E. 1967, MNRAS, 137, 337

\bibitem[{Fabricius {et~al.}(2021)Fabricius, Luri, Arenou, Babusiaux, Helmi,
  Muraveva, Reyl{\'e}, Spoto, Vallenari, Antoja, Balbinot, Barache, Bauchet,
  Bragaglia, Busonero, Cantat-Gaudin, Carrasco, Diakit{\'e}, Fabrizio,
  Figueras, Garc{\'\i}a-Gutierrez, Garofalo, Jordi, Kervella, Khanna, Leclerc,
  Licata, Lambert, Marrese, Masip, Ramos, Robichon, Robin, Romero-G{\'o}mez,
  Rubele, \& Weiler}]{Fabretal21a}
Fabricius, C., Luri, X., Arenou, F., {et~al.} 2021, A\&A, 649, A5

\bibitem[{Kuhn {et~al.}(2020)Kuhn, Hillenbrand, Carpenter, \&
  Avelar~Men{\'e}ndez}]{Kuhnetal20}
Kuhn, M.~A., Hillenbrand, L.~A., Carpenter, J.~M., \& Avelar~Men{\'e}ndez,
  {\'A}.~R. 2020, ApJ, 899, 128

\bibitem[{Lindegren {et~al.}(2021{\natexlab{a}})Lindegren, Bastian, Biermann,
  Bombrun, de~Torres, Gerlach, Geyer, Hern{\'a}ndez, Hilger, Hobbs, Klioner,
  Lammers, Ramos-Lerate, Steidelm{\"u}ller, Stephenson, \& van
  Leeuwen}]{Lindetal21b}
Lindegren, L., Bastian, U., Biermann, M., {et~al.} 2021{\natexlab{a}}, A\&A,
  649, A4

\bibitem[{Lindegren {et~al.}(2021{\natexlab{b}})Lindegren, $\, Hern{\'a}ndez,
  Bombrun, Ramos-Lerate, Steidelm{\"u}ller, Bastian, Biermann, de~Torres,
  Gerlach, Geyer, Hilger, Hobbs, Lammers, McMillan, Stephenson, Casta{\~n}eda,
  Davidson, Fabricius, Gracia-Abril, Portell, Rowell, Teyssier, Torra,
  Bartolom{\'e}, Clotet, Garralda, Gonz{\'a}lez-Vidal, Torra, Abbas, Altmann,
  Anglada~Varela, Balaguer-N{\'u}{\~n}ez, Balog, Barache, Becciani, Bernet,
  Bertone, Bianchi, Bouquillon, Brown, Bucciarelli, Busonero, Butkevich, Buzzi,
  Cancelliere, Carlucci, Charlot, Cioni, Crosta, Crowley, del Peloso, del Pozo,
  Drimmel, Esquej, Fienga, Fraile, Gai, Garc{\'\i}a-Reinaldos, Guerra, Hambly,
  Hauser, Jan{\ss}en, Jordan, Kostrzewa-Rutkowska, Lattanzi, Liao, Licata,
  Lister, L{\"o}ffler, Marchant, Masip, Mignard, Mints, Molina, Mora,
  Morbidelli, Murphy, Pagani, Panuzzo, Pe{\~n}alosa~Esteller, Poggio,
  Re~Fiorentin, Riva, Sagrist{\`a}~Sell{\'e}s, S{\'a}nchez~Gimenez, Sarasso,
  Sciacca, Siddiqui, Smart, Souami, Spagna, Steele, Taris, Utrilla, van Reeven,
  \& Vecchiato}]{Lindetal21a}
Lindegren, L., $\, \!$Klioner, S.~A., Hern{\'a}ndez, J., {et~al.}
  2021{\natexlab{b}}, A\&A, 649, A2

\bibitem[{Ma{\'\i}z~Apell{\'a}niz(2001)}]{Maiz01a}
Ma{\'\i}z~Apell{\'a}niz, J. 2001, AJ, 121, 2737

\bibitem[{Ma{\'\i}z~Apell{\'a}niz(2005)}]{Maiz05c}
Ma{\'\i}z~Apell{\'a}niz, J. 2005, in ESA Special Publication, Vol. 576, The
  Three-Dimensional Universe with Gaia, ed. C.~Turon, K.~S. O'Flaherty, \&
  M.~A.~C. Perryman, 179

\bibitem[{Ma{\'\i}z~Apell{\'a}niz(2017)}]{Maiz17a}
---. 2017, A\&A, 608, L8

\bibitem[{Ma{\'\i}z~Apell{\'a}niz(2019)}]{Maiz19}
---. 2019, A\&A, 630, A119 (Villafranca 0)

\bibitem[{Ma{\'\i}z~Apell{\'a}niz(2022)}]{Maiz22}
---. 2022, A\&A, 657, A130

\bibitem[{Ma{\'\i}z~Apell{\'a}niz {et~al.}(2008)Ma{\'\i}z~Apell{\'a}niz,
  Alfaro, \& Sota}]{Maizetal08a}
Ma{\'\i}z~Apell{\'a}niz, J., Alfaro, E.~J., \& Sota, A. 2008, arXiv:0804.2553

\bibitem[{Ma{\'\i}z~Apell{\'a}niz
  {et~al.}(2022{\natexlab{a}})Ma{\'\i}z~Apell{\'a}niz, Barb{\'a},
  Fern{\'a}ndez~Aranda, Pantaleoni~Gonz{\'a}lez, Crespo~Bellido, Sota, \&
  Alfaro}]{Maizetal22a}
Ma{\'\i}z~Apell{\'a}niz, J., Barb{\'a}, R.~H., Fern{\'a}ndez~Aranda, R.,
  {et~al.} 2022{\natexlab{a}}, A\&A, 657, A131 (Villafranca~II)

\bibitem[{Ma{\'\i}z~Apell{\'a}niz {et~al.}(2020)Ma{\'\i}z~Apell{\'a}niz,
  Crespo~Bellido, Barb{\'a}, Fern{\'a}ndez~Aranda, \& Sota}]{Maizetal20b}
Ma{\'\i}z~Apell{\'a}niz, J., Crespo~Bellido, P., Barb{\'a}, R.~H.,
  Fern{\'a}ndez~Aranda, R., \& Sota, A. 2020, A\&A, 643, A138 (Villafranca~I)

\bibitem[{Ma{\'\i}z~Apell{\'a}niz
  {et~al.}(2021{\natexlab{a}})Ma{\'\i}z~Apell{\'a}niz, Pantaleoni~Gonz{\'a}lez,
  \& Barb{\'a}}]{Maizetal21c}
Ma{\'\i}z~Apell{\'a}niz, J., Pantaleoni~Gonz{\'a}lez, M., \& Barb{\'a}, R.~H.
  2021{\natexlab{a}}, A\&A, 649, A13

\bibitem[{Ma{\'\i}z~Apell{\'a}niz
  {et~al.}(2022{\natexlab{b}})Ma{\'\i}z~Apell{\'a}niz, Pantaleoni~Gonz{\'a}lez,
  Barb{\'a}, \& Weiler}]{Maizetal22b}
Ma{\'\i}z~Apell{\'a}niz, J., Pantaleoni~Gonz{\'a}lez, M., Barb{\'a}, R.~H., \&
  Weiler, M. 2022{\natexlab{b}}, A\&A, 657, A72

\bibitem[{Ma{\'\i}z~Apell{\'a}niz \& Sota(2008)}]{MaizSota08}
Ma{\'\i}z~Apell{\'a}niz, J., \& Sota, A. 2008, in RMxAC, Vol.~33, 44--46

\bibitem[{Ma{\'\i}z~Apell{\'a}niz {et~al.}(2011)Ma{\'\i}z~Apell{\'a}niz, Sota,
  Walborn, Alfaro, Barb{\'a}, Morrell, Gamen, \& Arias}]{Maizetal11}
Ma{\'\i}z~Apell{\'a}niz, J., Sota, A., Walborn, N.~R., {et~al.} 2011, in HSA 6,
  467--472

\bibitem[{Ma{\'\i}z~Apell{\'a}niz {et~al.}(2019)Ma{\'\i}z~Apell{\'a}niz,
  Trigueros~P{\'a}ez, Jim{\'e}nez~Mart{\'\i}nez, Barb{\'a},
  Sim{\'o}n-D{\'\i}az, Pellerin, Negueruela, \& Souza~Le{\~a}o}]{Maizetal19a}
Ma{\'\i}z~Apell{\'a}niz, J., Trigueros~P{\'a}ez, E., Jim{\'e}nez~Mart{\'\i}nez,
  I., {et~al.} 2019, in HSA 10, 420 (\lili)

\bibitem[{Ma{\'\i}z~Apell{\'a}niz \& Weiler(2018)}]{MaizWeil18}
Ma{\'\i}z~Apell{\'a}niz, J., \& Weiler, M. 2018, A\&A, 619, A180

\bibitem[{Ma{\'\i}z~Apell{\'a}niz {et~al.}(2014)Ma{\'\i}z~Apell{\'a}niz, Evans,
  Barb{\'a}, Gr{\"a}fener, Bestenlehner, Crowther, Garc{\'\i}a, Herrero, Sana,
  Sim{\'o}n-D{\'\i}az, Taylor, van Loon, Vink, \& Walborn}]{Maizetal14a}
Ma{\'\i}z~Apell{\'a}niz, J., Evans, C.~J., Barb{\'a}, R.~H., {et~al.} 2014,
  A\&A, 564, A63

\bibitem[{Ma{\'\i}z~Apell{\'a}niz
  {et~al.}(2021{\natexlab{b}})Ma{\'\i}z~Apell{\'a}niz, Alfaro, Barb{\'a},
  Holgado, V{\'a}zquez-Rami{\'o}, Varela, Ederoclite, Lorenzo-Guti{\'e}rrez,
  Garc{\'\i}a-Lario, Garc{\'\i}a~Escudero, Garc{\'\i}a, \&
  Coelho}]{Maizetal21d}
Ma{\'\i}z~Apell{\'a}niz, J., Alfaro, E.~J., Barb{\'a}, R.~H., {et~al.}
  2021{\natexlab{b}}, MNRAS, 506, 3138

\bibitem[{Pantaleoni~Gonz{\'a}lez {et~al.}(2021)Pantaleoni~Gonz{\'a}lez,
  Ma{\'\i}z~Apell{\'a}niz, Barb{\'a}, \& Reed}]{Pantetal21}
Pantaleoni~Gonz{\'a}lez, M., Ma{\'\i}z~Apell{\'a}niz, J., Barb{\'a}, R.~H., \&
  Reed, B.~C. 2021, MNRAS, 504, 2968

\bibitem[{Reed(2003)}]{Reed03}
Reed, B.~C. 2003, AJ, 125, 2531

\bibitem[{Riello {et~al.}(2021)Riello, De~Angeli, Evans, Montegriffo, Carrasco,
  Busso, Palaversa, Burgess, Diener, Davidson, Rowell, Fabricius, Jordi,
  Bellazzini, Pancino, Harrison, Cacciari, van Leeuwen, Hambly, Hodgkin,
  Osborne, Altavilla, Barstow, Brown, Castellani, Cowell, De~Luise, Gilmore,
  Giuffrida, Hidalgo, Holland, Marinoni, Pagani, Piersimoni, Pulone, Ragaini,
  Rainer, Richards, Sanna, Walton, Weiler, \& Yoldas}]{Rieletal21}
Riello, M., De~Angeli, F., Evans, D.~W., {et~al.} 2021, A\&A, 649, A3

\end{thebibliography}

%
%
%
%

\end{document}